\documentclass[aps,reprint,pre,showpacs,longbibliography,floatfix]{revtex4-2}
\usepackage{times,amsmath,graphicx,latexsym}

\begin{document}

\title{Renormalized analytic solution for the enstrophy cascade in two-dimensional quantum turbulence}

\author{Andrew Forrester} 
\author{Han-Ching Chu} 
\author{Gary A. Williams}
\email{gaw@ucla.edu}
\affiliation{Department of Physics and Astronomy, University of California,
Los Angeles, CA 90095}
\date{\today}
\begin{abstract} 
The forward enstrophy cascade in two-dimensional quantum turbulence in a superfluid film connected to a thermal bath is investigated using a Fokker-Planck equation based on Kosterlitz-Thouless renormalization.  The steady-state cascade is formed by injecting vortex pairs of large initial separation at a constant rate.  They diffuse with a constant flux to smaller scales, finally annihilating when reaching the core size.   The energy spectrum varies as $k^{-3}$, similar to the spectrum known for 2D classical-fluid enstrophy cascades.  The dynamics of the cascade can also be studied, and for the case of a sharply peaked initial vortex-pair distribution, it takes about four eddy turnover times for the system to evolve to the decaying $k^{-3}$ cascade, in agreement with recent computer simulations.  These insights into the nature of the cascade also allow a better understanding of the phase-ordering process of temperature-quenched 2D superfluids, where the decay of the vorticity is found to proceed via the turbulent cascade.  This connection with turbulence may be a fundamental characteristic of phase-ordering in general.
 \end{abstract}  

\maketitle
\subsubsection{Introduction}
Constant-flux cascades have long played a role in understanding turbulence in fluids.  In classical 2D turbulence it is well known that there are two such cascades \cite{kraichnan80,tabeling,ecke}, an inverse cascade of energy to large length scales with a $k^{-5/3}$ (Kolmogorov) energy spectrum, and a forward cascade of enstrophy (vorticity) to small length scales, with a $k^{-3}$ energy spectrum (plus logarithmic corrections \cite{kraichnan_1971,falkovich,eyink2005,eyinkprl}).  There is no analytic theory of these cascades, and the primary evidence for their existence comes from numerical simulations of the Navier-Stokes equation \cite{boffetta2010}.  Since there is a wide range of length scales in the turbulence, it has long been recognized that renormalization group methods will be necessary to fully understand turbulent cascades in both two and three dimensions, and this has been a difficult program to implement \cite{zhou}, but recently there have been some advances in 2D methods \cite{Tarpin_2019}.

In 2D quantum turbulence in superfluids, with quantized point-like vortices, computer simulations have been able to identify the formation of an inverse energy cascade to large length scales, in which like-sign vortices begin to cluster together, a negative-temperature state \cite{anderson2013}.  More recently a computer simulation \cite{reeves} observed the development of an enstrophy cascade to small length scales with a $k^{-3}$ spectrum in decaying 2D quantum turbulence, validating the existence of the cascade, as proposed by us a number of years ago \cite{turb2001,forrester2014} involving the diffusive motion of vortex pairs of opposite circulation.   

Our system involves a 2D superfluid film on a substrate that acts as a thermal bath held at a relatively low temperature $T$, so there are almost no thermally excited vortices.  We continually inject at a constant rate vortex-antivortex pairs of large separation $R$ (out of equilibrium) , and due to the frictional forces on the vortex cores these pairs drift diffusively to smaller separation, finally annihilating when the separation becomes equal to the vortex core diameter $a_0$.  This constant flux of vorticity from large scales to small scales, with the rate of injection equal to the rate of annihilation and removal, constitutes the forward enstrophy cascade.  The frictional dissipation giving rise to the change in separation of the pairs plays a key role in the cascade, as it did in the computer simulation of Ref.[\cite{reeves}], where frictional dissipation was explicitly added to their Hamiltonian point-vortex model.  Energy is not conserved in the cascade; in our case any excess is simply absorbed by the thermal bath.

In this Rapid Communication we give a more detailed analytic solution of the cascade that utilizes a formulation of the 2D Kosterlitz-Thouless renormalization methods for the case of non-equilibrium vortex pairs \cite{ahns}.  The energy spectrum of the cascade is found to vary as $k^{-3}$, and the dynamics of the decay is found to be in agreement with the numerical simulations of Ref.\,\cite{reeves}.  We also highlight the relationship of the cascade to the dynamics of temperature-quenched 2D superfluids \cite{forrester2013}: the phase-ordering of the vortex decay in that case proceeds via the turbulent cascade.

\subsubsection{Theory} 
We consider an incompressible superfluid film connected to a thermal bath at low temperature $T =0.1\,T_{KT}$, where $T_{KT}$ is the critical Kosterlitz-Thouless temperature where thermally excited vortex pairs drive the superfluid density to zero.  Since the system is in contact with a thermal bath the negative-temperature states necessary for the inverse energy cascade cannot form, and so there is only the possibility of the forward enstrophy cascade.  The vortex dynamics are modeled by a Fokker-Planck equation \cite{ahns} describing the distribution of vortex pairs of separation $r$, with the addition of a forcing delta function to inject additional pairs of a fixed large separation $R$ into the film at a rate $\alpha$,
\begin{equation}
\frac{{\partial \,\Gamma }}{{\partial \,t}} =   \frac{1}{r}\,\:\frac{\partial }{{\partial r}}\left( {r\frac{{\partial \Gamma }}{{\,\partial r}} + 2\pi K\,\Gamma } \right) + {\alpha \;\delta^2 (\vec r - \vec R)}\quad.
\end{equation} 
This equation is set in dimensionless form with lengths in units of the vortex core diameter $a_0$, the vortex-pair distribution function $\Gamma$ in units $a_0^{4}$, and the time in units of the vortex diffusion time, $\tau_0 = a_0^{2} / 2D$ with $D$ the vortex diffusion coefficient arising from frictional forces on the vortex cores.  $K$ is the dimensionless superfluid density, $K = {{{\hbar ^2}{\sigma _s}}}/{{{m^2}{k_B}T}}$ with $\sigma_s$ the superfluid areal density and $m$ the atomic mass, and is determined from the Kosterlitz recursion relation \cite{kosterlitz}
\begin{equation}
\frac{{dK}}{{dr}} =  - 4{\pi ^3}{r^3}{K^2}\Gamma \quad.
\end{equation}
The term 
$2\pi K\Gamma$  in Eq.\,1 results from the attractive (screened) logarithmic interaction between pairs of opposite circulation. The 2D ``ring" delta function $\delta^2 ( \vec r - \vec R ) = \delta (r - R)/2\pi r$ injects pairs of separation R at random locations and orientations across the plane.  The dimensionless injection rate $\alpha$ is given by 
$\alpha  = a_0^2\dot Q\tau _0$ where $\dot Q$ is the number of vortex pairs of separation $R$ injected per unit area per time.  In the limit $\alpha  = 0$ and $\partial \Gamma / \partial t = 0$ these equations (1) and (2) reduce to the equilibrium renormalization equations of Kosterlitz and Thouless \cite{kosterlitz,kosterlitzrev}.  We note that Eq.\,(1) (without the delta function) was originally developed to understand the source of the dissipation found in thin $^4$He films  right at $T_{KT}$ in finite-frequency third sound \cite{rudnick1970,kotsubo} and torsional \cite{bishop} and shear oscillator \cite{wada} measurements.  At finite frequencies the friction on the vortex cores causes them to fall out of equilibrium with the applied flow, giving rise to dissipation and finite-size broadening, and Eq.\,(1) was found to accurately describe this situation.

The steady-state solutions ($\partial \Gamma / \partial t = 0$) found by iterating Eqs.\,(1) and (2) for the case R = 400 and $T=  0.1\,T_{KT}$ are shown  in Figure 1 for different values of $\alpha$, with the delta function at $R$ approximated with a strongly peaked Gaussian (width = 2), and extraction when the pairs annihilate at the core radius (and are absorbed by the thermal bath) is implemented with a boundary condition.   In the limit of low injection rates the vortex densities are well below the densities at $T_{KT}$, and the superfluid fraction is unaffected by the vortices, $K(r) = {K_0}$ where $K_0$ is the value at $r =1$ where $\sigma_s$ equals the ``bare" superfluid density $\sigma_s^0$.   Analytic solutions for the steady state can be found by adding a second delta function $-\alpha\,\delta^2 (\vec r - \vec 1)$ to the right-hand side of Eq.\,1 to account for the absorption of pairs at the same rate they are being injected, necessary to conserve vorticity.  This yields at low injection rates
\begin{equation}
\begin{array}{l}
\Gamma (r)  = {\Gamma _0} = \alpha /2\pi {K_0}\quad \quad (r < R)\\
\;\;\, = {\Gamma _0}{\left( {r/R} \right)^{ - 2\pi {K_0}}}\quad\quad\;\, (r > R).
\end{array}
\end{equation}  
For $r > R$ the solution is a quasi-thermal distribution extending from $R$, which arises from injected pairs initially at separation $R$ getting a thermal kick to higher separation.  

In the limit where the vortex density becomes comparable to that found at $T_{KT}$ (indicated by the dashed curve in Fig.\,1)  the superfluid fraction is rapidly driven to zero at a finite length scale $r_0$ that depends on the injection rate.  This effect on the superfluid density emphasizes  that the turbulent state is not just characterized by isolated dipole pairs, but in fact is a complex many-body state of smaller pairs screening the long-range interaction of larger pairs, and in this limit driving down the superfluid density.  If we approximate the drop in the superfluid density as a step function at $r_0$, analytic solutions can be found giving a logarithmic variation of $\Gamma$ at scales above $r_0$,
\begin{eqnarray}
\begin{array}{l}
\Gamma (r) = {\Gamma _0}\quad\quad\quad\quad\quad\quad(r < {r_0})\\
\quad  = {\Gamma _0}(1 + \ln (r/{r_0}))\quad ({r_0} < r < R)\\
\quad  = {\Gamma _0}(1 + \ln (R/{r_0}))\quad (r > R).
\end{array}
\end{eqnarray}
It can be seen that these analytic solutions match those found numerically in Fig.\,1.
The low bath temperature is the reason the KT recursion relations remain valid even when the superfluid density is zero, and there are then ``free" vortices of separation greater than R.  The equations do finally become unstable and blow up (similar to the equilibrium  behavior above $T_{KT}$) as the temperature is increased, depending on the value of $\alpha$; for 
$\alpha  = 1 \times {10^{ - 12}}$ this is around $T = 0.4\,T_{KT}$.

The steady-state diffusive flux of vortex pairs from Eq. 1 is 
$J =  - (r{{\partial \Gamma } \mathord{\left/
 {\vphantom {{\partial \Gamma } {\partial r + 2\pi K \Gamma)}}} \right.
 \kern-\nulldelimiterspace} {\partial r + 2\pi K \Gamma)}}$, and for both of the above solutions is a constant in the direction of small scales, $J = -\alpha$ for $r < R$, and zero for $r > R$.  
 
\begin{figure}[t]
\includegraphics[width=0.5\textwidth]{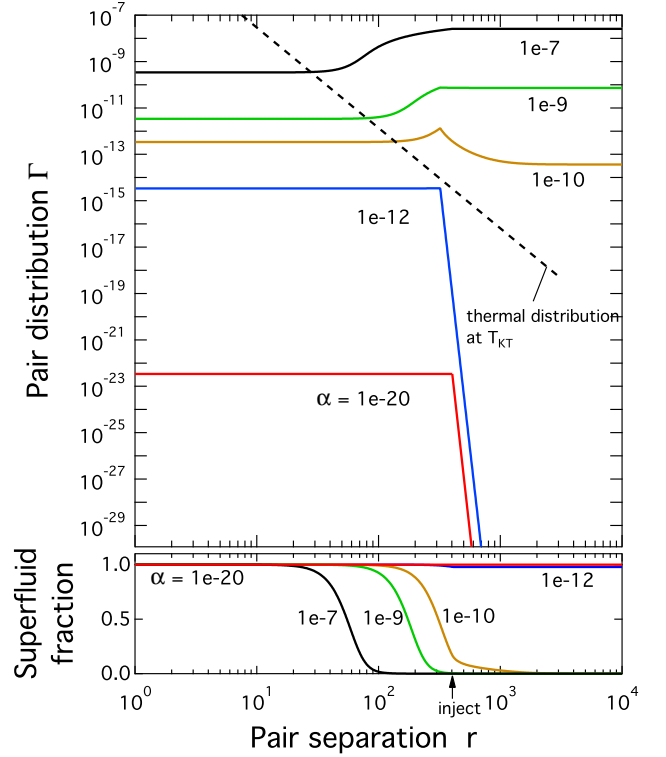}
\caption{Steady-state solutions of Eqs.[1] and [2] for the vortex-pair distribution function at different injection rates $\alpha$ at $R$ = 400, and $T =0.1\,T_{KT}$. }
\end{figure} 

 \begin{figure}[t]
\includegraphics[width=0.5\textwidth]{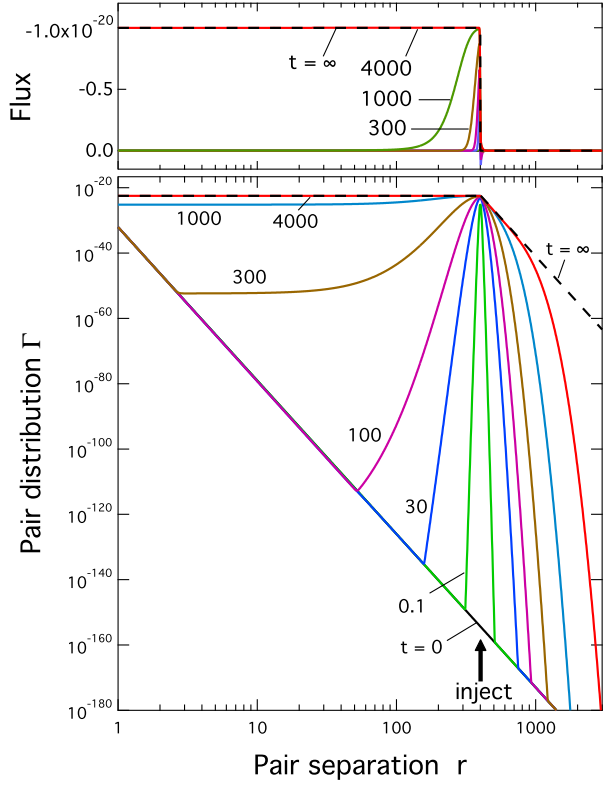} 
\caption{Growth of the cascade after turning on an injection rate $\alpha = 1\times10^{-20}$ at $R$ = 400, and $T =0.1\,T_{KT}$.}
\label{fig2}
\end{figure}

\subsubsection{Energy spectrum}
The  energy and enstrophy in the cascade region can be calculated following the results of Novikov \cite{novikov} and Tsubota \cite{tsubota2009}, where it is convenient now to return to dimensional units $r' = a_0r$, $R' = a_0R$, $\Gamma' = a_0^{-4}\,\Gamma$, etc.  The enstrophy per unit area in the cascade range for a uniform density $\rho_{pair}$ of vortex pairs of circulation $\kappa = 2\pi \hbar / m$ is then \cite{tsubota2009} 
\begin{equation}
\Omega = \int {\Omega (k')dk'}  = {\kappa}^2 \rho_{pair} \,\delta^2 ( \vec 0) 
\end{equation} 
where for our neutral gas of vortices ($N_+ = N_-$)
\begin{equation}
{\rho _{pair}} = \int {\Gamma '(r'){\kern 1pt} \,2\pi\,r'} dr'  
\end{equation}
and $\Omega(k')$ is the enstrophy spectrum.
Similarly, the real-space enstrophy injection rate is ${\kappa}^2\dot Q \,\delta^2 ( \vec 0)$, and the k-space transform is then $\eta =  {\kappa}^2\dot Q/a_0^2$.

The enstrophy spectrum can be found from a representation of the 2D delta function in terms of the Bessel function $J_0$,
\begin{equation}
{\delta ^2}(\vec 0) = \frac{1}{{{{\left( {2\pi } \right)}^2}}}\int {{e^{ - i\vec k' \cdot \vec r'}}{d^2}\vec k'}  = \frac{1}{{2\pi }}\int {{J_0}(k'r')k'dk'}
\end{equation}
and inserting this in Eq.(5) gives
\begin{equation}
\Omega (k') = \frac{{{\kappa ^2}k' \,}}{{2\pi}}\int {\Gamma '(r')} \;{J_0}(k'r')\,2\pi r'\,dr' \quad.
\end{equation}
The spectral kinetic energy $E(k')$ per unit mass is then given by the well-known relation with $\Omega (k')$ \cite{leith},
\begin{equation}
E(k') = \frac{{\Omega (k')}}{{{{\left( {k'} \right)}^2}}} = {{{\kappa ^2}}}{\left( {k'} \right)^{ - 3}}F
\end{equation}
where setting $z = k'r'$ the function $F$ is 
\begin{equation}
F = \int  {\Gamma ' (z/k')} \;{J_0}(z)\,z\,dz \quad.
\end{equation}
Evaluating $E(k')$ over the cascade range $a_0 < r' < R'$ of the constant solution of Eq.\,3 gives
\begin{equation}
E(k') = \frac{{\eta \,{\tau _0}}}{{{{2\pi }}\,{K_0}}}{(k')^{ - 3}}\tilde F
\end{equation}
with  
$\tilde F = k'R'J_1(k'R')-k'a_0'J_1(k'a_0)$.  Since the energy can only be positive definite, this rapid oscillation of the dimensionless $\tilde F$ is unphysical:  it is just the well-known result of imposing a square-wave cutoff on the transform over the finite length scale of the cascade region.  The energy using the solution of Eq.\,(4) is the same as (10), but with $\tilde F$ now a more complicated oscillating function.  The $k'^{-3}$ variation of Eq.\,(10) results only from the $const + ln(r)$ form found for $\Gamma$ in the solutions of Eqs.\,(3) and (4); any other variation would result in additional factors of $k'$ in the evaluation of Eq.\,(9). 
 \begin{figure}[t]
\includegraphics[width=0.5\textwidth]{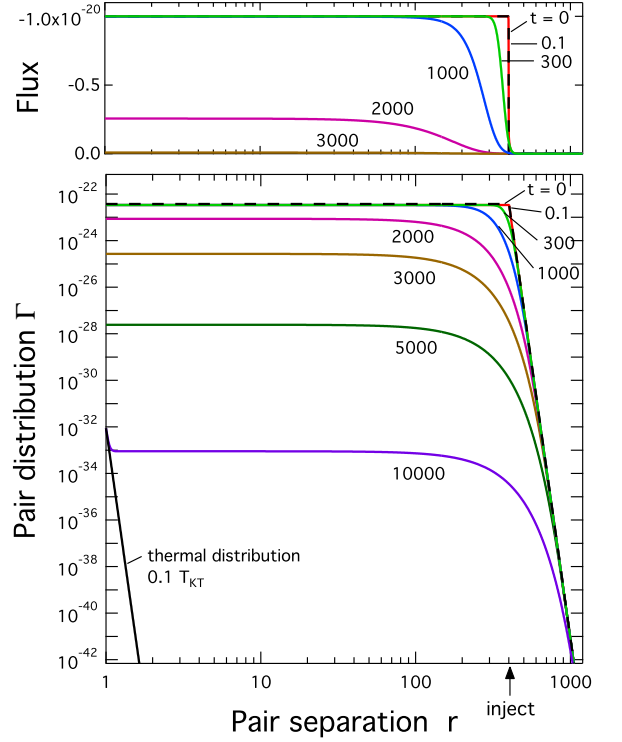} 
\caption{Decay of the vortex-pair density from the steady state reached in Fig.\,2, after switching off the injection.}
\label{fig3}
\end{figure}

  The linear variation of $E(k')$ with the enstrophy flux $\eta$ in Eq.\,(10) differs from the classical-fluid cascade, which varies as $\eta^{2/3}$.  The difference in our case is that $\tau_0$ is a new dimensional quantity that is the same for every vortex, and hence must appear as a factor in the energy.   $\tau_0$ is proportional to the frictional force on a vortex, and an increase in $\tau_0$ slows the motion of the vortices down the cascade, increasing the pair density, and hence increases the spectral energy.  This constrains the energy to be linear in $\eta$, since the product $\eta \tau_0$, with dimension $(time)^{-2}$, is the only dimensional possibility.  Similarly, the linear variation of the energy with temperature (from the factor $K_0$) comes about from the Einstein relation that lowers the mobility with increasing $T$, increasing the pair density and the energy.

\subsubsection{Dynamics of the cascade}  
The dynamics of the cascade can be studied by solving Eqs.\,(1) and (2) as a function of both time and separation using standard numerical techniques.  Figure 2 shows the growth of the distribution and the pair flux where a low injection rate $\alpha = 1\times10^{-20}$ is suddenly switched on at $t = 0$, and where again the delta function is approximated with a Gaussian of width 2 at $R = 400$ and an absorbing boundary condition is used at $r = 1$.  Initially the distribution just broadens as the pairs get thermal kicks to larger and smaller separations, but the frictional forces on the vortex cores also give rise to a net flux of pairs to smaller separations.  After a few hundred diffusion times the decaying pairs reach the core size separation where they begin to annihilate, and the energy is pulled out by the thermal bath.  This is basically the ``eddy turnover time" ${\tau _{eddy}} \approx 410\,{\tau _0}$ \cite{Rnote}.  After about 4 eddy times the distribution becomes relatively constant over $r < R$, which is the steady-state $k^{-3}$ cascade solution of Eq.\,(3).  The flux is initially only appreciable near the injection scale, and then finally reaches the constant value of $-\alpha$, the cascade state where the rate of pairs being injected at $R$ is equal to the rate of pairs being pulled out by the thermal bath at the scale $a_0$.  The total vortex density initially increases linearly with slope $\alpha$ before any pairs are pulled out at $a_0$, and then levels off to an equilibrium value once pairs begin annihilating.  
\begin{figure}[t]
\includegraphics[width=0.5\textwidth]{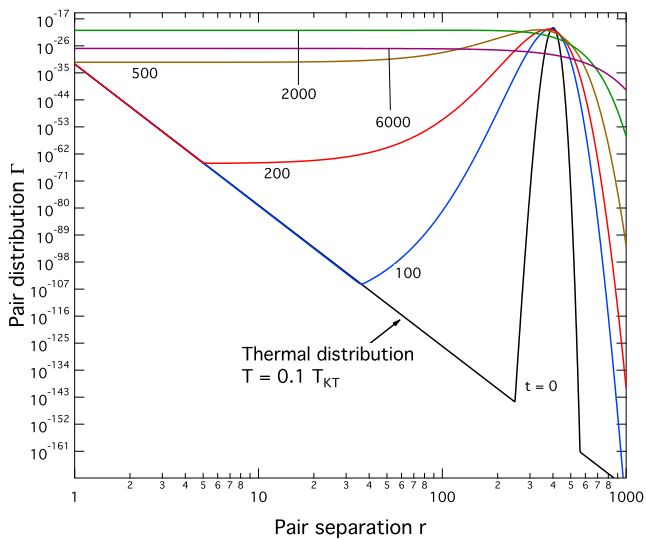} 
\caption{Decay of the distribution function from an initial spike at R = 400.}
\label{fig4}
\end{figure}

Upon switching off the injection, the decay of the cascade starts at the injection scale, as shown in Fig.\,3, since the flux of pairs away from $R$ is no longer being replenished by the injected pairs.    Once the region of diminished pairs begins to reach the scale $a_0$ the distribution starts to uniformly decrease, and in fact in this regime the solution becomes the exact solution found in Ref.\,\cite{forrester2013} for a quenched 2D superfluid, with a vortex pair density falling off as $t^{-(\pi K_0 - 1)}$.  We point out the fundamental relation between temperature-quenched superfluids and the vortex cascade, which both derive from the same equations.  In the temperature quench case the decay of the initial vortex distribution is entirely due to the development of the cascade flux that begins to remove the vortices by the annihilation at the smallest scale.  The phase-ordering dynamic length 
$\xi (t) = {\xi _0}{t^{ 1/z}}$, with $z = 2$ the dynamic exponent, is identified to be the growing extent of the flat region of the distribution function seen in Figs.\,1a and 3 of Ref.\cite{forrester2013}, defining the region of the $k^{-3}$ cascade where the small-scale vorticity is being removed \cite{davidson}.  

To more clearly illustrate the dynamics of freely decaying turbulence, we start with a spike in the distribution at $R = 400$ (a Gaussian of amplitude $1\times10^{-20}$ and width $\sigma = 20$), which otherwise is the thermal distribution at $T = 0.1\,T_{KT}$, and then monitor the subsequent time dependence, shown in Fig.\,4.  The spike relaxes again over about four eddy turnover times, where it evolves into the flat $k^{-3}$ cascade spectrum, and then uniformly continues to decay toward thermal equilibrium in exactly the same manner as the decay from the steady-state cascade or from the temperature quench.  This behavior is nearly identical to that seen in the simulations of Ref.\,\cite{reeves}, where a spike in the initial distribution (formed in their case by an actual real-space separation of the positive and negative circulation vortices) relaxes to the $k^{-3}$ spectrum over roughly four eddy times. Our turbulent cascade appears to be quite equivalent to the simulation results, although they find the energy spectrum proportional to $\eta^{2/3}$ instead of linear as in our case.  Since in the simulation frictional forces are not the same for every vortex (see Eq.\,(S5) of their Supplementary Material) there will not be a single diffusion time, and the only dimensional possibility is then $\eta^{2/3}$.  The vortices in the simulation start from a negative-temperature state, with like-sign vortices clustered together, but then rapidly diffuse together, mixing and increasing the entropy to what is likely an overall positive-temperature state, since annihilation has begun (which would not occur at negative temperatures).   There may still be negative-temperature regions remaining, since some regions of like-sign clustering are still found at late times.  Left unknown in this simulation is whether the system is actually a superfluid as the authors claim; based on our results in Fig.\,1 this seems very unlikely even at their lowest vortex densities.  We note the the pair distribution function could easily be computed from the vortex maps shown in Fig.\,1 of Ref.\,\cite{reeves}, using the pairing technique of Ref.\,\cite{cugliandolo}.

\subsubsection{Conclusions}
Experimentally, it is probably not likely that vortices in superfluid helium films can be easily accessed, but the situation in quasi-2D BEC systems may be more amenable.  Experiments have been able to identify the positions of both positive and negative circulation vortices \cite{shin}, and from such maps it should be possible to construct the pair distribution function as noted above.  It is now also possible to image magnetic vortices of both circulations in magnon-condensed ferromagnetic films \cite{radaelli}, and with suitable electrodes they can be injected at given rates into the films \cite{Zheng2017}.

In summary, we have constructed analytic solutions for the enstrophy cascade in two-dimensional quantum turbulence, using Kosterlitz renormalization.  The results are very simple and give a clear picture of the cascade as diffusing vortex pairs drifting to smaller separation under their mutual attraction, and annihilating at the smallest scale.  The energy spectrum varies as $k^{-3}$, quite similar to the classical-fluid case, and suggests that patch models of 2D Euler equations might be more generally useful \cite{melander}.  

There are also significant parallels with reaction-diffusion systems \cite{grindrod}.  The time decay characteristics of the enstrophy cascade allows an insight into the vortex decay found in thermally quenched 2D superfluids, where the phase-ordering proceeds via the constant-flux enstrophy cascade to small scales.  This connection with turbulent cascades may be a fundamental characteristic of the phase-ordering process of topological excitations in general.

\begin{acknowledgments}
This work was supported in part by a grant from the Julian Schwinger Foundation.  We thank George Morales and Seth Putterman for useful conversations.
\end{acknowledgments}


\section*{References}
\begin{thebibliography}{33}%
\makeatletter
\providecommand \@ifxundefined [1]{%
 \@ifx{#1\undefined}
}%
\providecommand \@ifnum [1]{%
 \ifnum #1\expandafter \@firstoftwo
 \else \expandafter \@secondoftwo
 \fi
}%
\providecommand \@ifx [1]{%
 \ifx #1\expandafter \@firstoftwo
 \else \expandafter \@secondoftwo
 \fi
}%
\providecommand \natexlab [1]{#1}%
\providecommand \enquote  [1]{``#1''}%
\providecommand \bibnamefont  [1]{#1}%
\providecommand \bibfnamefont [1]{#1}%
\providecommand \citenamefont [1]{#1}%
\providecommand \href@noop [0]{\@secondoftwo}%
\providecommand \href [0]{\begingroup \@sanitize@url \@href}%
\providecommand \@href[1]{\@@startlink{#1}\@@href}%
\providecommand \@@href[1]{\endgroup#1\@@endlink}%
\providecommand \@sanitize@url [0]{\catcode `\\12\catcode `\$12\catcode
  `\&12\catcode `\#12\catcode `\^12\catcode `\_12\catcode `\%12\relax}%
\providecommand \@@startlink[1]{}%
\providecommand \@@endlink[0]{}%
\providecommand \url  [0]{\begingroup\@sanitize@url \@url }%
\providecommand \@url [1]{\endgroup\@href {#1}{\urlprefix }}%
\providecommand \urlprefix  [0]{URL }%
\providecommand \Eprint [0]{\href }%
\providecommand \doibase [0]{https://doi.org/}%
\providecommand \selectlanguage [0]{\@gobble}%
\providecommand \bibinfo  [0]{\@secondoftwo}%
\providecommand \bibfield  [0]{\@secondoftwo}%
\providecommand \translation [1]{[#1]}%
\providecommand \BibitemOpen [0]{}%
\providecommand \bibitemStop [0]{}%
\providecommand \bibitemNoStop [0]{.\EOS\space}%
\providecommand \EOS [0]{\spacefactor3000\relax}%
\providecommand \BibitemShut  [1]{\csname bibitem#1\endcsname}%
\let\auto@bib@innerbib\@empty
\bibitem [{\citenamefont {Kraichnan}\ and\ \citenamefont
  {Montgomery}(1980)}]{kraichnan80}%
  \BibitemOpen
  \bibfield  {author} {\bibinfo {author} {\bibfnamefont {R.~H.}\ \bibnamefont
  {Kraichnan}}\ and\ \bibinfo {author} {\bibfnamefont {D.}~\bibnamefont
  {Montgomery}},\ }\bibfield  {title} {\bibinfo {title} {Two-dimensional
  turbulence},\ }\href@noop {} {\bibfield  {journal} {\bibinfo  {journal}
  {Reports on Prog. Phys.}\ }\textbf {\bibinfo {volume} {43}},\ \bibinfo
  {pages} {547} (\bibinfo {year} {1980})}\BibitemShut {NoStop}%
\bibitem [{\citenamefont {Tabeling}(2002)}]{tabeling}%
  \BibitemOpen
  \bibfield  {author} {\bibinfo {author} {\bibfnamefont {P.}~\bibnamefont
  {Tabeling}},\ }\bibfield  {title} {\bibinfo {title} {Two-dimensional
  turbulence},\ }\href@noop {} {\bibfield  {journal} {\bibinfo  {journal}
  {Phys. Reports}\ }\textbf {\bibinfo {volume} {362}},\ \bibinfo {pages} {1}
  (\bibinfo {year} {2002})}\BibitemShut {NoStop}%
\bibitem [{\citenamefont {Boffetta}\ and\ \citenamefont {Ecke}(2012)}]{ecke}%
  \BibitemOpen
  \bibfield  {author} {\bibinfo {author} {\bibfnamefont {G.}~\bibnamefont
  {Boffetta}}\ and\ \bibinfo {author} {\bibfnamefont {R.}~\bibnamefont
  {Ecke}},\ }\bibfield  {title} {\bibinfo {title} {Two-dimensional
  turbulence},\ }\href@noop {} {\bibfield  {journal} {\bibinfo  {journal} {Ann.
  Rev. Fluid Mech.}\ }\textbf {\bibinfo {volume} {44}},\ \bibinfo {pages} {427}
  (\bibinfo {year} {2012})}\BibitemShut {NoStop}%
\bibitem [{\citenamefont {Kraichnan}(1971)}]{kraichnan_1971}%
  \BibitemOpen
  \bibfield  {author} {\bibinfo {author} {\bibfnamefont {R.~H.}\ \bibnamefont
  {Kraichnan}},\ }\bibfield  {title} {\bibinfo {title} {Inertial-range transfer
  in two- and three-dimensional turbulence},\ }\href
  {https://doi.org/10.1017/S0022112071001216} {\bibfield  {journal} {\bibinfo
  {journal} {J Fluid Mech.}\ }\textbf {\bibinfo {volume} {47}},\ \bibinfo
  {pages} {525} (\bibinfo {year} {1971})}\BibitemShut {NoStop}%
\bibitem [{\citenamefont {Falkovich}\ and\ \citenamefont
  {Lebedev}(1994)}]{falkovich}%
  \BibitemOpen
  \bibfield  {author} {\bibinfo {author} {\bibfnamefont {G.}~\bibnamefont
  {Falkovich}}\ and\ \bibinfo {author} {\bibfnamefont {V.}~\bibnamefont
  {Lebedev}},\ }\bibfield  {title} {\bibinfo {title} {Universal direct cascade
  in two-dimensional turbulence},\ }\href@noop {} {\bibfield  {journal}
  {\bibinfo  {journal} {Phys. Rev. E}\ }\textbf {\bibinfo {volume} {50}},\
  \bibinfo {pages} {3883} (\bibinfo {year} {1994})}\BibitemShut {NoStop}%
\bibitem [{\citenamefont {Eyink}(2005)}]{eyink2005}%
  \BibitemOpen
  \bibfield  {author} {\bibinfo {author} {\bibfnamefont {G.~L.}\ \bibnamefont
  {Eyink}},\ }\bibfield  {title} {\bibinfo {title} {Locality of turbulent
  cascades},\ }\href
  {https://doi.org/https://doi.org/10.1016/j.physd.2005.05.018} {\bibfield
  {journal} {\bibinfo  {journal} {Physica D: Nonlinear Phenomena}\ }\textbf
  {\bibinfo {volume} {207}},\ \bibinfo {pages} {91 } (\bibinfo {year}
  {2005})}\BibitemShut {NoStop}%
\bibitem [{\citenamefont {Chen}\ \emph {et~al.}(2003)\citenamefont {Chen},
  \citenamefont {Ecke}, \citenamefont {Eyink}, \citenamefont {Wang},\ and\
  \citenamefont {Zuoli}}]{eyinkprl}%
  \BibitemOpen
  \bibfield  {author} {\bibinfo {author} {\bibfnamefont {C.}~\bibnamefont
  {Chen}}, \bibinfo {author} {\bibfnamefont {R.}~\bibnamefont {Ecke}}, \bibinfo
  {author} {\bibfnamefont {G.}~\bibnamefont {Eyink}}, \bibinfo {author}
  {\bibfnamefont {X.}~\bibnamefont {Wang}},\ and\ \bibinfo {author}
  {\bibfnamefont {X.}~\bibnamefont {Zuoli}},\ }\bibfield  {title} {\bibinfo
  {title} {Physical mechanism of the two-dimensional enstrophy cascade},\
  }\href@noop {} {\bibfield  {journal} {\bibinfo  {journal} {Phys. Rev. Lett.}\
  }\textbf {\bibinfo {volume} {91}},\ \bibinfo {pages} {214501} (\bibinfo
  {year} {2003})}\BibitemShut {NoStop}%
\bibitem [{\citenamefont {Boffetta}\ and\ \citenamefont
  {Musacchio}(2010)}]{boffetta2010}%
  \BibitemOpen
  \bibfield  {author} {\bibinfo {author} {\bibfnamefont {G.}~\bibnamefont
  {Boffetta}}\ and\ \bibinfo {author} {\bibfnamefont {S.}~\bibnamefont
  {Musacchio}},\ }\bibfield  {title} {\bibinfo {title} {Evidence for the double
  cascade scenario in two-dimensional turbulence},\ }\href
  {https://doi.org/10.1103/PhysRevE.82.016307} {\bibfield  {journal} {\bibinfo
  {journal} {Phys. Rev. E}\ }\textbf {\bibinfo {volume} {82}},\ \bibinfo
  {pages} {016307} (\bibinfo {year} {2010})}\BibitemShut {NoStop}%
\bibitem [{\citenamefont {Zhou}(2010)}]{zhou}%
  \BibitemOpen
  \bibfield  {author} {\bibinfo {author} {\bibfnamefont {Y.}~\bibnamefont
  {Zhou}},\ }\bibfield  {title} {\bibinfo {title} {Renormalization group theory
  for fluid and plasma turbulence},\ }\href
  {https://doi.org/https://doi.org/10.1016/j.physrep.2009.04.004} {\bibfield
  {journal} {\bibinfo  {journal} {Phys. Reports}\ }\textbf {\bibinfo {volume}
  {488}},\ \bibinfo {pages} {1 } (\bibinfo {year} {2010})}\BibitemShut
  {NoStop}%
\bibitem [{\citenamefont {Tarpin}\ \emph {et~al.}(2019)\citenamefont {Tarpin},
  \citenamefont {Canet}, \citenamefont {Pagani},\ and\ \citenamefont
  {Wschebor}}]{Tarpin_2019}%
  \BibitemOpen
  \bibfield  {author} {\bibinfo {author} {\bibfnamefont {M.}~\bibnamefont
  {Tarpin}}, \bibinfo {author} {\bibfnamefont {L.}~\bibnamefont {Canet}},
  \bibinfo {author} {\bibfnamefont {C.}~\bibnamefont {Pagani}},\ and\ \bibinfo
  {author} {\bibfnamefont {N.}~\bibnamefont {Wschebor}},\ }\bibfield  {title}
  {\bibinfo {title} {Stationary, isotropic and homogeneous two-dimensional
  turbulence: a first non-perturbative renormalization group approach},\ }\href
  {https://doi.org/10.1088/1751-8121/aaf3f0} {\bibfield  {journal} {\bibinfo
  {journal} {J. of Phys. A: Math. and Theor.}\ }\textbf {\bibinfo {volume}
  {52}},\ \bibinfo {pages} {085501} (\bibinfo {year} {2019})}\BibitemShut
  {NoStop}%
\bibitem [{\citenamefont {Reeves}\ \emph {et~al.}(2013)\citenamefont {Reeves},
  \citenamefont {Billam}, \citenamefont {Anderson},\ and\ \citenamefont
  {Bradley}}]{anderson2013}%
  \BibitemOpen
  \bibfield  {author} {\bibinfo {author} {\bibfnamefont {M.~T.}\ \bibnamefont
  {Reeves}}, \bibinfo {author} {\bibfnamefont {T.~P.}\ \bibnamefont {Billam}},
  \bibinfo {author} {\bibfnamefont {B.~P.}\ \bibnamefont {Anderson}},\ and\
  \bibinfo {author} {\bibfnamefont {A.~S.}\ \bibnamefont {Bradley}},\
  }\bibfield  {title} {\bibinfo {title} {Inverse energy cascade in forced
  two-dimensional quantum turbulence},\ }\href@noop {} {\bibfield  {journal}
  {\bibinfo  {journal} {Phys. Rev. Lett.}\ }\textbf {\bibinfo {volume} {110}},\
  \bibinfo {pages} {104501} (\bibinfo {year} {2013})}\BibitemShut {NoStop}%
\bibitem [{\citenamefont {Reeves}\ \emph {et~al.}(2017)\citenamefont {Reeves},
  \citenamefont {Billam}, \citenamefont {Yu},\ and\ \citenamefont
  {Bradley}}]{reeves}%
  \BibitemOpen
  \bibfield  {author} {\bibinfo {author} {\bibfnamefont {M.~T.}\ \bibnamefont
  {Reeves}}, \bibinfo {author} {\bibfnamefont {T.~P.}\ \bibnamefont {Billam}},
  \bibinfo {author} {\bibfnamefont {X.}~\bibnamefont {Yu}},\ and\ \bibinfo
  {author} {\bibfnamefont {A.~S.}\ \bibnamefont {Bradley}},\ }\bibfield
  {title} {\bibinfo {title} {Enstrophy cascade in decaying two-dimensional
  quantum turbulence},\ }\href {https://doi.org/10.1103/PhysRevLett.119.184502}
  {\bibfield  {journal} {\bibinfo  {journal} {Phys. Rev. Lett.}\ }\textbf
  {\bibinfo {volume} {119}},\ \bibinfo {pages} {184502} (\bibinfo {year}
  {2017})}\BibitemShut {NoStop}%
\bibitem [{\citenamefont {Chu}\ and\ \citenamefont
  {Williams}(2001)}]{turb2001}%
  \BibitemOpen
  \bibfield  {author} {\bibinfo {author} {\bibfnamefont {H.-C.}\ \bibnamefont
  {Chu}}\ and\ \bibinfo {author} {\bibfnamefont {G.~A.}\ \bibnamefont
  {Williams}},\ }\bibfield  {title} {\bibinfo {title} {Nonequilibrium vortex
  dynamics in superfluid phase transitions and superfluid turbulence},\ }in\
  \href@noop {} {\emph {\bibinfo {booktitle} {Quantized Vortex Dynamics and
  Superfluid Turbulence}}},\ \bibinfo {series and number} {\bibinfo {series}
  {Lecture Notes in Physics}\ No.\ \bibinfo {number} {571}},\ \bibinfo {editor}
  {edited by\ \bibinfo {editor} {\bibfnamefont {C.~F.}\ \bibnamefont
  {Barenghi}}, \bibinfo {editor} {\bibfnamefont {R.~J.}\ \bibnamefont
  {Donnelly}},\ and\ \bibinfo {editor} {\bibfnamefont {W.~F.}\ \bibnamefont
  {Vinen}}}\ (\bibinfo  {publisher} {Springer-Verlag},\ \bibinfo {address}
  {Heidelberg},\ \bibinfo {year} {2001})\ pp.\ \bibinfo {pages}
  {226--232}\BibitemShut {NoStop}%
\bibitem [{\citenamefont {Forrester}\ and\ \citenamefont
  {Williams}(2014)}]{forrester2014}%
  \BibitemOpen
  \bibfield  {author} {\bibinfo {author} {\bibfnamefont {A.}~\bibnamefont
  {Forrester}}\ and\ \bibinfo {author} {\bibfnamefont {G.~A.}\ \bibnamefont
  {Williams}},\ }\bibfield  {title} {\bibinfo {title} {Dynamics of the forward
  vortex cascade in two-dimensional quantum turbulence},\ }\href@noop {}
  {\bibfield  {journal} {\bibinfo  {journal} {J. Phys.: Conf. Ser.}\ }\textbf
  {\bibinfo {volume} {568}},\ \bibinfo {pages} {012031} (\bibinfo {year}
  {2014})}\BibitemShut {NoStop}%
\bibitem [{\citenamefont {Ambegaokar}\ \emph {et~al.}(1980)\citenamefont
  {Ambegaokar}, \citenamefont {Halperin}, \citenamefont {Nelson},\ and\
  \citenamefont {Siggia}}]{ahns}%
  \BibitemOpen
  \bibfield  {author} {\bibinfo {author} {\bibfnamefont {V.}~\bibnamefont
  {Ambegaokar}}, \bibinfo {author} {\bibfnamefont {B.}~\bibnamefont
  {Halperin}}, \bibinfo {author} {\bibfnamefont {D.}~\bibnamefont {Nelson}},\
  and\ \bibinfo {author} {\bibfnamefont {E.}~\bibnamefont {Siggia}},\
  }\bibfield  {title} {\bibinfo {title} {Dynamics of superfluid films},\
  }\href@noop {} {\bibfield  {journal} {\bibinfo  {journal} {Phys. Rev. B}\
  }\textbf {\bibinfo {volume} {21}},\ \bibinfo {pages} {1806} (\bibinfo {year}
  {1980})}\BibitemShut {NoStop}%
\bibitem [{\citenamefont {Forrester}\ \emph {et~al.}(2013)\citenamefont
  {Forrester}, \citenamefont {Chu},\ and\ \citenamefont
  {Williams}}]{forrester2013}%
  \BibitemOpen
  \bibfield  {author} {\bibinfo {author} {\bibfnamefont {A.}~\bibnamefont
  {Forrester}}, \bibinfo {author} {\bibfnamefont {H.-C.}\ \bibnamefont {Chu}},\
  and\ \bibinfo {author} {\bibfnamefont {G.~A.}\ \bibnamefont {Williams}},\
  }\bibfield  {title} {\bibinfo {title} {Exact solution for vortex dynamics in
  temperature quenches of two-dimensional superfluids},\ }\href@noop {}
  {\bibfield  {journal} {\bibinfo  {journal} {Phys. Rev. Lett.}\ }\textbf
  {\bibinfo {volume} {110}},\ \bibinfo {pages} {165303} (\bibinfo {year}
  {2013})}\BibitemShut {NoStop}%
\bibitem [{\citenamefont {Kosterlitz}(1974)}]{kosterlitz}%
  \BibitemOpen
  \bibfield  {author} {\bibinfo {author} {\bibfnamefont {J.~M.}\ \bibnamefont
  {Kosterlitz}},\ }\bibfield  {title} {\bibinfo {title} {The critical
  properties of the two-dimensional xy model},\ }\href@noop {} {\bibfield
  {journal} {\bibinfo  {journal} {J. Phys. C}\ }\textbf {\bibinfo {volume}
  {7}},\ \bibinfo {pages} {1046} (\bibinfo {year} {1974})}\BibitemShut
  {NoStop}%
\bibitem [{\citenamefont {Kosterlitz}(2016)}]{kosterlitzrev}%
  \BibitemOpen
  \bibfield  {author} {\bibinfo {author} {\bibfnamefont {J.~M.}\ \bibnamefont
  {Kosterlitz}},\ }\bibfield  {title} {\bibinfo {title} {Kosterlitz--thouless
  physics: a review of key issues},\ }\href@noop {} {\bibfield  {journal}
  {\bibinfo  {journal} {Rep. Prog. Phys.}\ }\textbf {\bibinfo {volume} {79}},\
  \bibinfo {pages} {026001} (\bibinfo {year} {2016})}\BibitemShut {NoStop}%
\bibitem [{\citenamefont {Rudnick}\ and\ \citenamefont
  {Fraser}(1970)}]{rudnick1970}%
  \BibitemOpen
  \bibfield  {author} {\bibinfo {author} {\bibfnamefont {I.}~\bibnamefont
  {Rudnick}}\ and\ \bibinfo {author} {\bibfnamefont {J.~C.}\ \bibnamefont
  {Fraser}},\ }\bibfield  {title} {\bibinfo {title} {Third sound and the
  superfluid parameters of thin helium films},\ }\href@noop {} {\bibfield
  {journal} {\bibinfo  {journal} {J. Low Temp. Phys.}\ }\textbf {\bibinfo
  {volume} {3}},\ \bibinfo {pages} {225} (\bibinfo {year} {1970})}\BibitemShut
  {NoStop}%
\bibitem [{\citenamefont {Kotsubo}\ and\ \citenamefont
  {Williams}(1986)}]{kotsubo}%
  \BibitemOpen
  \bibfield  {author} {\bibinfo {author} {\bibfnamefont {V.}~\bibnamefont
  {Kotsubo}}\ and\ \bibinfo {author} {\bibfnamefont {G.~A.}\ \bibnamefont
  {Williams}},\ }\bibfield  {title} {\bibinfo {title} {Superfluid transition of
  $^{4}\mathrm{He}$ films adsorbed in porous materials},\ }\href@noop {}
  {\bibfield  {journal} {\bibinfo  {journal} {Phys. Rev. B}\ }\textbf {\bibinfo
  {volume} {33}},\ \bibinfo {pages} {6106} (\bibinfo {year}
  {1986})}\BibitemShut {NoStop}%
\bibitem [{\citenamefont {Bishop}\ and\ \citenamefont {Reppy}(1978)}]{bishop}%
  \BibitemOpen
  \bibfield  {author} {\bibinfo {author} {\bibfnamefont {D.~J.}\ \bibnamefont
  {Bishop}}\ and\ \bibinfo {author} {\bibfnamefont {J.~D.}\ \bibnamefont
  {Reppy}},\ }\bibfield  {title} {\bibinfo {title} {Study of the superfluid
  transition in two-dimensional $^4$he films},\ }\href@noop {} {\bibfield
  {journal} {\bibinfo  {journal} {Phys. Rev. Lett.}\ }\textbf {\bibinfo
  {volume} {40}},\ \bibinfo {pages} {1727} (\bibinfo {year}
  {1978})}\BibitemShut {NoStop}%
\bibitem [{\citenamefont {Hieda}\ \emph {et~al.}(2009)\citenamefont {Hieda},
  \citenamefont {Matsuda}, \citenamefont {Kato}, \citenamefont {Matsushita},\
  and\ \citenamefont {Wada}}]{wada}%
  \BibitemOpen
  \bibfield  {author} {\bibinfo {author} {\bibfnamefont {M.}~\bibnamefont
  {Hieda}}, \bibinfo {author} {\bibfnamefont {K.}~\bibnamefont {Matsuda}},
  \bibinfo {author} {\bibfnamefont {T.}~\bibnamefont {Kato}}, \bibinfo {author}
  {\bibfnamefont {T.}~\bibnamefont {Matsushita}},\ and\ \bibinfo {author}
  {\bibfnamefont {N.}~\bibnamefont {Wada}},\ }\bibfield  {title} {\bibinfo
  {title} {Extremely high frequency dependence of two-dimensional superfluid
  onset},\ }\href@noop {} {\bibfield  {journal} {\bibinfo  {journal} {J. Phys.
  Soc. Japan}\ }\textbf {\bibinfo {volume} {78}},\ \bibinfo {pages} {033604}
  (\bibinfo {year} {2009})}\BibitemShut {NoStop}%
\bibitem [{\citenamefont {Novikov}(1976)}]{novikov}%
  \BibitemOpen
  \bibfield  {author} {\bibinfo {author} {\bibfnamefont {E.~A.}\ \bibnamefont
  {Novikov}},\ }\bibfield  {title} {\bibinfo {title} {Dynamics and statics of a
  system of vortices},\ }\href@noop {} {\bibfield  {journal} {\bibinfo
  {journal} {Sov. Phys. JETP}\ }\textbf {\bibinfo {volume} {41}},\ \bibinfo
  {pages} {937} (\bibinfo {year} {1976})}\BibitemShut {NoStop}%
\bibitem [{\citenamefont {Numasato}\ and\ \citenamefont
  {Tsubota}(2010)}]{tsubota2009}%
  \BibitemOpen
  \bibfield  {author} {\bibinfo {author} {\bibfnamefont {R.}~\bibnamefont
  {Numasato}}\ and\ \bibinfo {author} {\bibfnamefont {M.}~\bibnamefont
  {Tsubota}},\ }\bibfield  {title} {\bibinfo {title} {Possibility of inverse
  energy cascade in two-dimensional quantum turbulence},\ }\href@noop {}
  {\bibfield  {journal} {\bibinfo  {journal} {J. Low Temp. Phys.}\ }\textbf
  {\bibinfo {volume} {158}},\ \bibinfo {pages} {415} (\bibinfo {year}
  {2010})}\BibitemShut {NoStop}%
\bibitem [{\citenamefont {Leith}(1968)}]{leith}%
  \BibitemOpen
  \bibfield  {author} {\bibinfo {author} {\bibfnamefont {C.~E.}\ \bibnamefont
  {Leith}},\ }\bibfield  {title} {\bibinfo {title} {Diffusion approximation for
  two-dimensional turbulence},\ }\href@noop {} {\bibfield  {journal} {\bibinfo
  {journal} {Phys. Fluids}\ }\textbf {\bibinfo {volume} {11}},\ \bibinfo
  {pages} {671} (\bibinfo {year} {1968})}\BibitemShut {NoStop}%
\bibitem [{Rno()}]{Rnote}%
  \BibitemOpen
  \href@noop {} {\bibinfo  {journal} {By varying $R$ we have numerically
  confirmed that $\tau_{eddy}$ varies as $R^2$, since $\tau_{eddy} \sim
  \rho_{pair}/4 \dot Q \propto R^2$}\ }\BibitemShut {NoStop}%
\bibitem [{\citenamefont {Lowe}\ and\ \citenamefont
  {Davidson}(2005)}]{davidson}%
  \BibitemOpen
\bibfield  {journal} {  }\bibfield  {author} {\bibinfo {author} {\bibfnamefont
  {A.}~\bibnamefont {Lowe}}\ and\ \bibinfo {author} {\bibfnamefont
  {P.}~\bibnamefont {Davidson}},\ }\bibfield  {title} {\bibinfo {title} {The
  evolution of freely-decaying, isotropic, two-dimensional turbulence},\ }\href
  {https://doi.org/https://doi.org/10.1016/j.euromechflu.2004.09.003}
  {\bibfield  {journal} {\bibinfo  {journal} {European Journal of Mechanics -
  B/Fluids}\ }\textbf {\bibinfo {volume} {24}},\ \bibinfo {pages} {314 }
  (\bibinfo {year} {2005})},\ \bibinfo {note} {. It is interesting in this
  article that a $t^{1/2}$ growing length scale is also found for the freely
  decaying enstrophy cascade in simulations of classical fluids.}\BibitemShut
  {Stop}%
\bibitem [{\citenamefont {Jelic}\ and\ \citenamefont
  {Cugliandolo}(2011)}]{cugliandolo}%
  \BibitemOpen
  \bibfield  {author} {\bibinfo {author} {\bibfnamefont {A.}~\bibnamefont
  {Jelic}}\ and\ \bibinfo {author} {\bibfnamefont {L.~F.}\ \bibnamefont
  {Cugliandolo}},\ }\bibfield  {title} {\bibinfo {title} {Quench dynamics of
  the 2d xy model},\ }\href@noop {} {\bibfield  {journal} {\bibinfo  {journal}
  {J. Stat. Mech.}\ ,\ \bibinfo {pages} {P02032}} (\bibinfo {year}
  {2011})}\BibitemShut {NoStop}%
\bibitem [{\citenamefont {Seo}\ \emph {et~al.}(2017)\citenamefont {Seo},
  \citenamefont {Ko}, \citenamefont {Kim},\ and\ \citenamefont {Shin}}]{shin}%
  \BibitemOpen
  \bibfield  {author} {\bibinfo {author} {\bibfnamefont {S.}~\bibnamefont
  {Seo}}, \bibinfo {author} {\bibfnamefont {B.}~\bibnamefont {Ko}}, \bibinfo
  {author} {\bibfnamefont {J.~H.}\ \bibnamefont {Kim}},\ and\ \bibinfo {author}
  {\bibfnamefont {Y.}~\bibnamefont {Shin}},\ }\bibfield  {title} {\bibinfo
  {title} {Observation of vortex-antivortex pairing in decaying 2d turbulence
  of a superfluid gas},\ }\href@noop {} {\bibfield  {journal} {\bibinfo
  {journal} {Scientific Reports}\ }\textbf {\bibinfo {volume} {7}},\ \bibinfo
  {pages} {4587} (\bibinfo {year} {2017})}\BibitemShut {NoStop}%
\bibitem [{\citenamefont {Chmiel}\ \emph {et~al.}(2018)\citenamefont {Chmiel},
  \citenamefont {Waterfield~Price}, \citenamefont {Johnson}, \citenamefont
  {Lamirand}, \citenamefont {Schad}, \citenamefont {van~der Laan},
  \citenamefont {Harris}, \citenamefont {Irwin}, \citenamefont {Rzchowski},
  \citenamefont {Eom},\ and\ \citenamefont {Radaelli}}]{radaelli}%
  \BibitemOpen
  \bibfield  {author} {\bibinfo {author} {\bibfnamefont {F.~P.}\ \bibnamefont
  {Chmiel}}, \bibinfo {author} {\bibfnamefont {N.}~\bibnamefont
  {Waterfield~Price}}, \bibinfo {author} {\bibfnamefont {R.~D.}\ \bibnamefont
  {Johnson}}, \bibinfo {author} {\bibfnamefont {A.~D.}\ \bibnamefont
  {Lamirand}}, \bibinfo {author} {\bibfnamefont {J.}~\bibnamefont {Schad}},
  \bibinfo {author} {\bibfnamefont {G.}~\bibnamefont {van~der Laan}}, \bibinfo
  {author} {\bibfnamefont {D.~T.}\ \bibnamefont {Harris}}, \bibinfo {author}
  {\bibfnamefont {J.}~\bibnamefont {Irwin}}, \bibinfo {author} {\bibfnamefont
  {M.~S.}\ \bibnamefont {Rzchowski}}, \bibinfo {author} {\bibfnamefont {C.~B.}\
  \bibnamefont {Eom}},\ and\ \bibinfo {author} {\bibfnamefont {P.~G.}\
  \bibnamefont {Radaelli}},\ }\bibfield  {title} {\bibinfo {title} {Observation
  of magnetic vortex pairs at room temperature in a planar alpha-fe2o3/co
  heterostructure},\ }\href@noop {} {\bibfield  {journal} {\bibinfo  {journal}
  {Nature Materials}\ }\textbf {\bibinfo {volume} {17}},\ \bibinfo {pages}
  {581} (\bibinfo {year} {2018})}\BibitemShut {NoStop}%
\bibitem [{\citenamefont {Zheng}\ and\ \citenamefont {Chen}(2017)}]{Zheng2017}%
  \BibitemOpen
  \bibfield  {author} {\bibinfo {author} {\bibfnamefont {Y.}~\bibnamefont
  {Zheng}}\ and\ \bibinfo {author} {\bibfnamefont {W.~J.}\ \bibnamefont
  {Chen}},\ }\bibfield  {title} {\bibinfo {title} {Characteristics and
  controllability of vortices in ferromagnetics, ferroelectrics, and
  multiferroics},\ }\href@noop {} {\bibfield  {journal} {\bibinfo  {journal}
  {Reports on Prog. Phys.}\ }\textbf {\bibinfo {volume} {80}},\ \bibinfo
  {pages} {086501} (\bibinfo {year} {2017})}\BibitemShut {NoStop}%
\bibitem [{\citenamefont {Melander}\ \emph {et~al.}(1986)\citenamefont
  {Melander}, \citenamefont {Zabusky},\ and\ \citenamefont
  {Styczek}}]{melander}%
  \BibitemOpen
  \bibfield  {author} {\bibinfo {author} {\bibfnamefont {M.~V.}\ \bibnamefont
  {Melander}}, \bibinfo {author} {\bibfnamefont {N.~J.}\ \bibnamefont
  {Zabusky}},\ and\ \bibinfo {author} {\bibfnamefont {A.~S.}\ \bibnamefont
  {Styczek}},\ }\bibfield  {title} {\bibinfo {title} {A moment model for vortex
  interactions of two-dimensional euler equations},\ }\href@noop {} {\bibfield
  {journal} {\bibinfo  {journal} {J. Fluid Mech.}\ }\textbf {\bibinfo {volume}
  {167}},\ \bibinfo {pages} {95} (\bibinfo {year} {1986})}\BibitemShut
  {NoStop}%
\bibitem [{\citenamefont {Grindrod}(1991)}]{grindrod}%
  \BibitemOpen
  \bibfield  {author} {\bibinfo {author} {\bibfnamefont {P.}~\bibnamefont
  {Grindrod}},\ }\href@noop {} {\emph {\bibinfo {title} {Patterns and Waves:
  The Theory and Applications of Reaction-Diffusion Equations}}}\ (\bibinfo
  {publisher} {Clarendon Press},\ \bibinfo {address} {Oxford},\ \bibinfo {year}
  {1991})\BibitemShut {NoStop}%
\end{thebibliography}
\end{document}